\documentclass[twocolumn,showpacs]{revtex4-1}
\usepackage{amssymb}
\usepackage{amsmath}
\usepackage{graphicx}
\usepackage{color}

\newcommand{\nn}{\nonumber}

\begin{document}

\title{Time evolution of entropy in a growth model: Dependence on the description}

\author{Segun \surname{Goh}}
\affiliation{Center for Theoretical Physics, Seoul National University, Seoul 08826, Korea}
\author{Jungzae \surname{Choi}}
\affiliation{Department of Physics and Department of Chemical Engineering, Keimyung University, Daegu 42601, Korea}
\author{M.Y. \surname{Choi}}
\affiliation{Department of Physics and Astronomy and Center for Theoretical Physics, Seoul National University, Seoul 08826, Korea}
\author{Byung-Gook \surname{Yoon}}
\thanks{E-mail: bgyoon@ulsan.ac.kr}
\affiliation{Department of Physics, University of Ulsan, Ulsan 44610, Korea}

\begin{abstract}
Entropy plays a key role in statistical physics of complex systems, which in general exhibit diverse aspects of emergence on different scales. However, it still remains not fully resolved how entropy varies with the coarse-graining level and the description scale. In this paper, we consider a Yule-type growth model, where each element is characterized by its size being either continuous or discrete. Entropy is then defined directly from the probability distribution of the states of all elements as well as from the size distribution of the system. Probing in detail their relations and time evolutions, we find that heterogeneity in addition to correlations between elements could induce loss of information during the coarse-graining procedure. It is also revealed that the expansion of the size space domain depends on the description level, leading to a difference between the continuous description and the discrete one.
\end{abstract}


\maketitle

\section{Introduction}

Undoubtedly, entropy is one of the most important quantity in physics~\cite{Wehrl1978, Sethna2006}. It connects the thermodynamic behavior of a macroscopic system with the configurations of microscopic states~\cite{Landau1980}, giving rise to modern statistical physics. In particular, entropy, incorporated with information after the seminal work on information theory~\cite{Shannon2015}, plays a central role in physics of complex systems~\cite{Costa2002, Gellmann2004, Anand2009}. For instance, information exchange dynamics was proposed as the underlying mechanism of self-organized criticality~\cite{Choi2005, Kim2013}; the maximum entropy model was proposed to understand the physics of biological systems such as species abundance~\cite{Frank2011} and the collective behavior in neural networks~\cite{Schneidman2006, Tkavcik2014}.

Notwithstanding the fundamental and practical importance as mentioned above, some properties of entropy still remain somewhat controversial. Specifically, the second law of thermodynamics, which states the nondecreasing time evolution of entropy of an isolated system, is still an actively studied topic~\cite{Deffner2013, Parrondo2015, Seifert2016}. According to the fluctuation theorem in particular~\cite{Crooks1999, Collin2005}, the nondecreasing property of entropy is feasible only on the macroscopic scale while a decrease of entropy may indeed be observed in a small system.

Here it should be noted that entropy may not be singly defined across the coarse-graining level or the scale of description.
As a representative example, one may consider a system consisting of many elements, and define the entropy from the probability for the system to be in given configuration, i.e., for each element to be in given state, or from the state distribution of all the elements in the system.
Henceforth, for convenience, we call the former entropy (defined by the probability of the system configuration) `fine-grained entropy' and the latter one (by the state distribution function of elements) `coarse-grained entropy'.
The general master equation governing the time evolution of the probability allows one to probe the time evolution of the entropy as well. To clarify the difference between the two entropies, we analyze the simple growth model, with no production or with uniform size production~\cite{Choi2009, Goh2010}.
The state of each element is specified by its `size', which can in general take continuous values. Nevertheless the description based on discrete values of the size can also be adopted; both the continuous and the discrete descriptions are examined.
This growth model bears skew distributions, as manifested by the time evolution obtained from the master equation, and thus provide a good framework to probe the issues mentioned above.

Naively, one may expect that the coarse-grained entropy is equivalent to the fine-grained entropy if the elements are independent of each other. In such a case, the whole system can be decomposed fully into single elements and the coarse-graining procedure should not introduce information loss. Examining the system with uniform size production, however, we find that the independence between elements is not sufficient: heterogeneity of elements can serve as an additional source of the information loss in the coarse-graining procedure.
Further, resolution of the description is also proved to play a significant role. The expansion rate of the size space domain in the continuous description is qualitatively different from that in the discrete description and as a consequence, a term describing such a growing space domain is introduced in the continuous description of the system.

Meanwhile, the evolution equations for the probability density functions of both systems are analytically tractable as shown in Ref.~\cite{Goh2010}, at least after a sufficiently long time. We may thus use the known analytic expression of entropy for those cases. On the other hand, the time evolution at earlier stages can in general be obtained only via numerical methods. Moreover, a system with more complex growth mechanism resists analytical treatment, compelling one to resort to numerical calculation. Therefore, we perform extensive numerical simulations as well.

This paper consists of five sections: In Sec. II, we formulate the entropy dynamics of the system governed by a general master equation. Section III describes the time evolution of entropy applied to the growth model, while numerical results are presented in Sec. IV. Finally, a brief summary is given in Sec. V.

\section{Time evolution of entropy}
\label{sec_numerics}

We consider a system of $N$ elements, the $i$th of which is characterized by its size $x_i$ ($i = 1, \ldots, N$). The configuration of the system is specified by the sizes of all elements, $\{x_1^{}, \dots, x_N^{}\}$ or shortly by $\{x_i\}$. If $x_i$ is a continuous variable, the probability density $P(\{x_i\}; t)$ for the system to be in configuration $\{x_i\}$ at time $t$ is governed by the master equation
\begin{widetext}
  \begin{equation}
    \label{eq: master equation}
     \frac {d} {dt}  P(\{x_i\};t) = \sum_{i=1}^{N} \int d x_i'\,
	  \left[ \omega(x_i' \rightarrow x_i^{}) P(x_1^{},\dots,x_i',\dots,x_{N}^{};t) - \omega(x_i \rightarrow x_i') P(\{x_i\};t) \right] ,
  \end{equation}
\end{widetext}
where $\omega(x_i \rightarrow x_i')$ is the transition rate for the $i$th element to change its size from $x_i$ to $x_i'$. We are also interested in the size distribution $f(x,t)$, related to the probability density $P(\{x_i\}; t)$ via
\begin{equation}
  \label{eq: size distribution}
  f(x,t) = \frac{1}{N} \int d^N x \sum_{i=1}^{N} \delta(x_i - x) P(\{x_i\}; t)\,,
\end{equation}
where $\int d^N\!x \equiv \int d x_{1}^{} \cdots \int d x_{N}^{}$.

Using the probability density and the size distribution, one can define entropy in two ways and probe the time evolution of the two: the fine-grained entropy and the coarse-grained entropy, the relation between which is of interest here.
First, we define the fine-grained entropy to be
\begin{equation}
S_P(t)\equiv -\int d^N x\, P(\{x_i \};t) \ln{P(\{x_i\};t)},
\end{equation}
the time evolution of which is obtained from Eq.~\eqref{eq: master equation}:
\begin{widetext}
\begin{align}
\label{eq: dSdt general}
\frac{d S_P(t)}{dt} &=-\int d^N x\left[ 1+\ln{P(\{x_i\};t)} \right]\frac{d}{d t}P(\{x_i\};t) \nn \\
&=-\int d^N x \left[ 1+\ln{P(\{x_i\};t)} \right] \sum_{i=1}^N \int dx_i^\prime
 \left[ \omega (x_i^\prime \to x_i)P(x_1, \ldots, x_i^\prime, \ldots, x_N;t)-\omega (x_i \to x_i^\prime)P(\{x_i\};t) \right] \nn \\
&=-\int d^N x \sum_{i=1}^N \int dx_i^\prime
\left[ \omega (x_i \to x_i^\prime ) P(\{x_i\};t) \ln{p(x_1, \ldots, x_i^\prime, \ldots, x_N;t)}
-\omega (x_i \to x_i^\prime ) P(\{x_i\};t) \ln{P(\{x_i\};t)} \right] \nn \\
&=-\left\langle \sum_{i=1}^N \int dx_i^\prime \omega (x_i \to x_i^\prime )\ln{\frac{P(x_1, \ldots, x_i^\prime, \ldots, x_N;t)}{P(\{x_i\};t)}} \right\rangle
\end{align}
\end{widetext}
with $\displaystyle \langle \mathcal{O} \rangle \equiv \int d^N x\, \mathcal{O} P(\{x_i\};t)$. This equation is generally applicable to the differential entropy of the system governed by the master equation.
It is well known that the differential entropy of the system described by continuous variables suffers from the divergence of the information capacity. Here we deal with the time evolution of the (differential) entropy, where such divergence cancels out. Accordingly, there does not occur the problem of divergence.

We next define the coarse-grained (differential) entropy 
according to
\begin{equation}
\label{eq: diff ent}
S_f(t) \equiv -\int dx\, f(x,t) \ln{f(x,t)},
\end{equation}
which yields
\begin{equation}
\frac{d S_f}{d t}=-\int dx \frac{\partial f(x,t)}{\partial t}\left[ 1+\ln{f(x,t)} \right].
\end{equation}
Equation~\eqref{eq: size distribution} allows one to rewrite $S_f$ in the form:
\begin{equation}
S_f(t) =\left\langle\frac{1}{N}\sum_{i=1}^N  \ln{f(x_i, t)} \right\rangle ,
\end{equation}
which is just the Lyapunov exponent~\cite{Eckmann1985} of the mapping
\begin{equation}
x_{n+1} =F(x_n)
\end{equation}
with $F(x)$ defined by $dF(x)/dx \equiv f(x)$.
It is thus manifested that the differential entropy is the Lyapunov exponent or the dynamic entropy with the index $n$ counting the time step.
Note here that the probability density function quantifies the relation between adjacent stochastic variables $x_n$ and $x_{n+1}$. If the correlations between them are small, the actual trajectory of the series $\{x_i\}$ should be rather unstable, characterized by sensitivity to the initial condition $x_1$ and accordingly by a large value of the Lyapunov exponent. Indeed, entropy is a representative measure for regularity of the system, and it is natural to interpret the differential entropy as the dynamic entropy of the stochastic process defined through the probability density function.

We can also extend the analysis to the system whose number of elements varies in time. In this case, the entropy evolves in time as follows:
\begin{widetext}
\begin{align}
\label{eq: num non conserved ent}
\frac{d S_P}{dt} &\equiv \lim_{\Delta t \to 0} \frac{S_P (t{+}\Delta t)- S_P (t)}{\Delta t} \nn \\
&= \lim_{\Delta t \to 0}\frac{1}{\Delta t} \left[ \int d^{N{+}\Delta N} x\, P(x_1, \ldots, x_{N{+}\Delta N}; t{+}\Delta t)
	\left[ \ln{P(\{x_i\}; t{+}\Delta t)} \right. \right. \nn \\
		&\quad ~~\left.\left. + \ln{P(x_{N{+}1}, \ldots, x_{N{+}\Delta N}; t{+}\Delta t |x_1,\ldots,x_N; t)} \right]
 -\int d^N x\, P(\{x_i\};t) \ln{P(\{x_i\};t)} \right] \nn \\
&=\frac{d S_P^{(0)}}{dt}
	+\frac{dN}{dt}\frac{1}{\Delta N}\int d^{N{+}\Delta N}x\, P(x_{1}, \ldots, x_{N{+}\Delta N}; t+\Delta t) \ln{P(x_{N{+}1},\ldots, x_{N{+}\Delta N}; t{+}\Delta t| \{x_i\}; t)} \nn \\
&\equiv \frac{d S_P^{(0)}}{dt} + \frac{dN}{dt}s^{(1)},
\end{align}
\end{widetext}
where the first term in the last line represents the time evolution of the number conserving part $S_P^{(0)}$ and the second term corresponds to the production of conditional entropy $s^{(1)}$ (per element) associated with the birth of new elements.

It is straightforward to apply this formulation to a system with discrete size variables: Replacing the integration $\int d^N x$ and the delta function $\delta (x_i -x)$ by the summation $\sum_{x_i}$ and the Kronecker delta $\delta_{x_i, x}$, respectively, one can easily obtain the evolution equation for the information entropy (instead of the differential entropy) in a similar form. In addition, we here point out that $S_P$ and $S_f$ are in general not equivalent. In the system of elements coupled with each other, the entropy of the system is not extensive and to replace the system configuration probability $P(\{x_i\}; t)$ by the coarse-grained state distribution function $f(x,t)$ would cause information loss arising from the ignorance of correlations between elements.

\section{Independent elements: Application to Growth model}

In the case that the elements of a system are independent of each other, we have $P(\{x_i\};t)=P_1(x_1;t)\cdots P_N(x_N;t)$ where $P_i(x;t)$ is the probability for the size of the $i$th element to be $x$. The fine-grained entropy is then given by the sum
\begin{align}
S_P(t) &= -\int d^N x \, \prod_i P_i (x_i;t) \ln{\left( \prod_i P_i (x_i;t) \right)} \nn \\
&= -\sum_i \int dx\, P_i (x;t) \ln{P_i (x;t)} ,
\label{SP}
\end{align}
where extensiveness is obvious.
However, such independence between elements does not guarantee the equivalence between the fine-grained entropy and the coarse-grained one. Since Eq.~\eqref{eq: size distribution} reduces to
\begin{equation}
f(x,t) =\frac{1}{N}\sum_i P_i (x;t),
\end{equation}
the coarse-grained entropy reads
\begin{equation}
\label{eq: S neq Sf}
S_f(t)=-\int dx \, \frac{1}{N}\sum_{i=1}^N P_i(x;t) \ln{\left[ \frac{1}{N}\sum_{i=1}^N P_i(x;t) \right]}.
\end{equation}
Comparison of Eqs. (\ref{SP}) and (\ref{eq: S neq Sf}) indeed shows that $S_P$ and $S_f$ are not necessarily equivalent: In fact the Cauchy-Schwarz inequality indicates that $NS_f \leq S_P$. Here the equality requires additional assumption that every element has the same probability for the size: $P_1(x;t)=\cdots=P_N(x;t)\equiv P(x;t)$. This gives $f(x, t) = P(x; t)$ and the equality $S_P = NS_f$.
Accordingly, in the case of a heterogeneous system, the coarse-grained entropy is larger than the fine-grained entropy; this reflects the loss of information in the coarse-graining procedure, arising from the disregard of details of the element sizes.
Equipped with these observations, we now consider the growth model and probe the time evolution of entropy in various cases.

\subsection{Simple growth without production}
For convenience, we begin with a brief summary of the growth model developed and analyzed in Refs.~\cite{Choi2009, Goh2010}. First, we consider the system whose number of elements is fixed. In this case of a number conserving system without production, the only process involved is the size change (growth) by the amount proportional to the current size and the transition rate takes the form
\begin{equation}
  \label{eq: transition rate}
  \omega(x_i \rightarrow x_i') = \lambda \delta[x_i'-(1{+}b)x_i]
\end{equation}
with the (mean) growth rate $\lambda$ and the growth factor $b$. Making use of Eq.~(\ref{eq: master equation}), we obtain the evolution equation for the size distribution $f(x,t)$:
\begin{equation}
  \label{eq: number conserved}
  \frac{\partial{}}{\partial{t}} f(x,t) = - \lambda f(x,t) + \frac{\lambda}{1+b} f \left( \frac{x}{1+b},t \right).
\end{equation}
It is known that the log-normal distribution of the form
\begin{equation}
  \label{eq: log-normal distribution}
  f(x, t) =\frac{1}{\sqrt{2 \pi} \sigma_t^{}x} \exp \left[-\frac{ \left(\ln x -\mu_t^{} \right) ^2}{2 \sigma_t^2} \right]
\end{equation}
provides an asymptotic solution of Eq. (\ref{eq: number conserved}). Specifically, under the initial condition $f(x,0)=\delta (x-1)$, we have the mean $\mu_t = \lambda t \ln{(1{+}b)}$ and the deviation $\sigma_t = \sqrt{\lambda t} \ln{(1{+}b)}$ \cite{Goh2014}.

We also probe the system in the discrete description, where the transition rate reads
\begin{equation}
  \label{eq: transition rate disc}
  \omega(x_i \rightarrow x_i') = \lambda \delta_{x_i',(1{+}b)x_i}.
\end{equation}
This in turn leads to the evolution equation in a slightly modified form
\begin{equation}
  \label{eq: number conserved disc}
  \frac{\partial{}}{\partial{t}} f(x,t) = - \lambda f(x,t) + \lambda f \left( \frac{x}{1+b},t \right).
\end{equation}
When the initial size of every element is given by unity, the size at a later time can take only the discrete value $x = (1+b)^k$ for some integer $k$. We thus write simply $f\left((1{+}b)^k; t\right)\equiv p_k(t)$, which evolves in time according to
\begin{equation}
  \label{eq: p_k}
    \frac{\partial{}}{\partial{t}} p_k (t) = - \lambda  p_k (t) + \lambda p_{k-1} (t).
\end{equation}
It is easy to obtain the solution of Eq.~(\ref{eq: p_k}):
\begin{equation}
  \label{eq: poisson}
	p_k (t) = \frac{1}{k!}( \lambda t)^k e^{-\lambda t},
\end{equation}
which is the Poisson distribution~\cite{Risken1996}.
Note that the normalization condition is now given by
\begin{equation}
\sum_k f\left((1{+}b)^k ,t\right) = 1.
\end{equation}

In both continuous and discrete descriptions, elements grow independently of each other and the probability is the same for every element, leading to the relation $P(\{x_i\};t)= P_1(x_1;t) \cdots P_N(x_N;t) \equiv [P(x;t)]^N$. In consequence, the time evolution is simplified to take the form
\begin{align}
\frac{\partial S_P}{\partial t} &= -N \lambda \left\langle \ln{\frac{P\left((1{+}b)x;t\right)}{P(x;t)}} \right\rangle \\
\frac{\partial S_f}{\partial t} &= -\lambda \left\langle \ln{\frac{f\left((1{+}b)x,t\right)}{f(x,t)}} \right\rangle,
\label{eq: number conserved entropy}
\end{align}
where, along with $f(x, t)=P(x;t)$, the only difference is the factor $N$ representing the extensive property.
Henceforth, one can safely probe the time evolution using $f(x,t)$ instead of $P(\{x_i \};t)$. This approach is not applicable to the system in which couplings between elements may not be neglected. Note also that the above relations are valid for the discrete description as well, with the integration in the averaging procedure replaced by the summation. However, the entropy in the continuous description and that in the discrete one could be different, as they are governed by different time evolution equations, Eqs.~\eqref{eq: number conserved} and \eqref{eq: number conserved disc}.

We now use the solution of the time evolution equation to pursue specifically the time evolution of the entropy. Inserting the log-normal distribution to Eq.~\eqref{eq: number conserved entropy}, we obtain the asymptotic behavior of the entropy in the form
\begin{equation}
\frac{d S_f}{dt}=\lambda \ln{(1+b)}+\frac{1}{2t}.
\end{equation}
One can also compute the entropy directly from Eq.~\eqref{eq: diff ent}, and obtain the consistent result
\begin{equation}
\label{eq: num conserved ent asympt}
S_f(t) = \lambda t \ln{(1+b)} -\frac{1}{2}\ln{t } -\frac{1}{2}\lambda + C_C,
\end{equation}
where $C_C$ is a constant.

In the discrete description, entropy can be computed from the Poisson distribution, similarly to the continuous one. The entropy for the Poisson distribution is well known and behaves asymptotically as~\cite{Evans1988}
\begin{equation}
\label{eq: num conserved ent asympt disc}
S_f \approx \frac{1}{2}\ln{\lambda t} +C_D,
\end{equation}
where $C_D$ is a constant depending on the growth rate $\lambda$. Note that the asymptotic behavior is free of the growth factor $b$ as expected. Note also that the main difference between the continuous description and the discrete description is given by the term $\lambda t\ln{(1{+}b)}$, which is a direct consequence of the $1/x$ factor in the log-normal distribution~\cite{Goh2014}. In deriving the log-normal distribution from the Gaussian distribution, the factor $1/x$ is brought by the change of the measure $dX = dx/x$ in the logarithmic transformation $x\to X\equiv \ln{x}$. Therefore, we conclude that the term has its origin solely in the growing domain of the size space in the continuous description. We will return to this issue in Sec. IV [see Eq.~\eqref{eq: ent dk vs ds}].

\subsection{Growth with production of new elements}
Next, we consider the case that the total number of elements varies with time, i.e., $N=N(t)$, and each element tends to produce a new one with rate $r$ (thus the total number of elements increases in proportion to the current number: $\dot{N} = rN$). The time evolution equation for the size distribution obtains the form:
\begin{equation}
  \label{eq: f evolution 2}
  \frac{\partial{f(x,t)}}{\partial{t}} = -(r+\lambda) f(x,t) + \frac{\lambda}{1+b} f \left({\frac{x}{1+b},t} \right) + rg(x,t),
\end{equation}
where $g(x,t)$ is the size distribution function of newly produced elements.
In this work, we deal with the case that new elements are produced in uniform size $x_0$, i.e., $g(x,t) = \delta (x-x_0)$. The stationary distribution is then given by a power-law function for $x >x_0$ \cite{Goh2010}:
\begin{equation}
 \label{eq: stationary power}
 f (x) \sim x^{-\alpha}
\end{equation}
with the exponent
\begin{equation}
\label{eq: exponent}
\alpha = 1 + \frac{\ln{(1+r/\lambda)}}{\ln{(1+b)}}.
\end{equation}
In the discrete description, the evolution equation for $p_k$ under uniform size production, corresponding to Eq.~(\ref{eq: f evolution 2}), reads
\begin{equation}
  \label{eq: p_k u}
    \frac{\partial{}}{\partial{t}} p_k (t) = - (r+\lambda)  p_k (t) + \lambda p_{k-1} (t) + r \delta_{k,0},
\end{equation}
of which the exact stationary solution is given by
\begin{equation}
\label{eq: stationary discrete}
p_k =\frac{r}{r+\lambda} (1+b)^{-k\ln{(1+r/\lambda)}/\ln{(1+b)}}.
\end{equation}

When new elements of uniform size are produced, the entropy can still be decomposed into the entropy component of each element. On the other hand, the relation $f(x,t)=P(x;t)$ is not satisfied because the entropy of an element produced at time $t_1$ and that at $t_2$ are obviously different from each other. It is therefore expected that the fine-grained entropy $S_P(t)$ and the coarse-grained entropy $S_f(t)$ are not equivalent in this case; this will be confirmed by computing the stationary entropy values specifically. Further, to circumvent the extensiveness of the fine-grained entropy growing with the number of elements, we focus on the entropy per element $s \equiv S_P /N$ rather than $S_P$. Of course, this is not the case for the coarse-grained entropy $S_f$.

We first probe the stationary value of $s$ which should be computed directly from Eq.~\eqref{eq: num non conserved ent}. The time evolution is governed by
\begin{equation}
\frac{ds}{dt}=-rs+\frac{ds^{(0)}}{dt}+rs^{(1)},
\end{equation}
where $s^{(0)}$ is the entropy per element in the number conserving system with the asymptotic behavior given by Eq.~\eqref{eq: num conserved ent asympt}. The additional term $rs^{(1)}$ originates from the last term in Eq.~\eqref{eq: num non conserved ent}. Note that in this model system, $P(x_{N{+}1},\ldots, x_{N{+}\Delta N}; t{+}\Delta t|x_1,\ldots, x_N; t)=P(x_{N{+}1},\ldots, x_{N{+}\Delta N}; t{+}\Delta t)$ and accordingly, $s^{(1)}$ is simply the entropy per new element. If we further assume $s^{(0)} = s(0)$, the time evolution of entropy per element is described by:
\begin{equation}
\label{eq: ent sol uni prod}
s(t)=e^{-rt} s(0)+r\int_0^t dt' e^{-r(t-t')} s(t-t').
\end{equation}
Therefore, if we know the single-element entropy in the number conserving system, we can precisely compute the entropy in the uniform production case. Fortunately, we have $s(t)=S_f(t)$ and also obtained the time evolution of $S_f (t)$ in Eq.~\eqref{eq: num conserved ent asympt}. Neglecting the first term on the right hand side of Eq.~\eqref{eq: ent sol uni prod}, we approximate the stationary value of the entropy as follows:
\begin{align}
s &\approx r\lim_{t \to \infty} \int_0^t dt^\prime e^{-r(t-t^\prime)}s(t{-}t^\prime) \nn \\
&\approx -\lim_{t\to 0} \lambda \ln{(1{+}b)} \frac{rt+1}{r} e^{-rt} +\frac{\lambda}{r}\ln{(1{+}b)} \nn \\
&\quad +\frac{1}{2}\lim_{t\to\infty}\left[ {\rm Ei}(-rt') - e^{-rt'} \ln{(\lambda t')} \right]_{t'=0}^{t'=t} \nn \\
&= \frac{\lambda}{r}\ln{(1{+}b)}-\frac{1}{2}\ln{r} -\frac{1}{2}\ln{\lambda} + B_1 ,
\label{eq: fine h stationary}
\end{align}
where Ei is the exponential integral~\cite{Abramowitz1964}
and $B_1$ is a constant. Similarly, in the discrete description, the stationary value obtains
\begin{equation}
\label{eq: fine h stationary disc}
s = -\frac{1}{2}\ln{r} -\frac{1}{2}\ln{\lambda} + B_2 .
\end{equation}
Even though the constant shift $B_2$ remains unclarified, the dependency of the entropy on the model parameters $\lambda$, $b$, and $r$ is fully specified.

We then turn to the coarse-grained entropy $S_f$. In the continuous description, we can compute the stationary value from the exact form of $f(x,t)$ [see Eq.~\eqref{eq: stationary power}]. Performing the integration, we thus obtain the coarse-grained entropy 
\begin{equation}
\label{eq: coarsen h stationary}
S_f =-\ln{(\alpha -1)}+\frac{\alpha}{\alpha-1}.
\end{equation}
In the discrete description, on the other hand, we can exactly compute the stationary value of entropy from Eq.~\eqref{eq: stationary discrete}, to obtain the form
\begin{align}
S_f &= -\sum_{n=0}^\infty \frac{r}{r+\lambda} (1+b)^{-n\ln{(1{+}r/\lambda)}/\ln{(1{+}b)}} \nn \\ &~~\times \ln{\left[ \frac{r}{r+\lambda} (1+b)^{-n\ln{(1{+}r/\lambda)}/\ln{(1{+}b)}} \right]}.
\end{align}
Denoting $A \equiv (1+b)^{-\ln{(1{+}r/\lambda)}/\ln{(1{+}b)}}= \lambda (r+\lambda)^{-1}$, we obtain
\begin{align}
\label{eq: coarsen h stationary disc}
S_f &= -\frac{r}{r+\lambda}\ln{\frac{r}{r+\lambda}}\sum_{n=0}^\infty A^n  - \frac{r}{r+\lambda}\ln{A}\sum_{n=0}^\infty n A^n \nn \\
& = -\ln{\left( \frac{r}{r+\lambda} \right)} -\frac{\lambda}{r}\ln{\left( \frac{\lambda}{r+\lambda} \right)}.
\end{align}

We now ponder on the mechanism for the emergence of the stationary power-law distribution. If there is no production, the system evolves to the disordered state as the entropy increases indefinitely. In the presence of uniform size production, on the other hand, the state of newly produced elements is fully ordered in the sense that the additional entropy contributions from the new elements vanish. As a result of appropriate mixing of these two components, there emerges a stationary state whose asymptotic entropy is finite. For this stationary state, we have confirmed that both $s$ and $S_f$ are finite.

Finally, from Eqs.~\eqref{eq: fine h stationary} and~\eqref{eq: coarsen h stationary} [or from Eqs.~\eqref{eq: fine h stationary disc} and~\eqref{eq: coarsen h stationary disc}], it is evident that $S_f\neq s$ even with the constant shift disregarded. As we could not specify the initial value of the entropy, it is still unclear whether the coarse-grained entropy is larger than the fine-grained one due to the information loss in the coarse-graining procedure. However, the dependency on the model parameters is clearly distinguished and we conclude that the coarse-grained entropy could differ from the fine-grained one even in the case of a non-interacting system. The issue associated with the information loss will be clarified by the numerical results in the next section.

\section{Numerical Simulations}

Let us first describe briefly the algorithm to compute the time evolution of (differential) entropy. The procedure begins with the numerical integration of the evolution equation to obtain the distribution function as a function of time (and $x$). In the numerical integration, we use the fourth-order Runge-Kutta method for time integration with the time step $\delta t=0.01$, while dividing the positive $x$ space into segments of equal or variable length(s). Finally, we calculate the differential entropy in the continuous description at each time, which is defined to be
\begin{equation}
  \label{eq: differ}
  S_f(t) = - \int_0^{\infty} dx f(x,t) \ln f(x,t) = - \left< \ln f(x,t) \right>.
\end{equation}
For numerical integration, we use the simplest approximation for Eq.~(\ref{eq: differ}):
\begin{equation}
  \label{eq: differ 2}
  S_f(t) = - \sum_x  f(x,t) \ln f(x,t) \delta x .
\end{equation}
In the discrete description, the entropy is computed directly from the definition of the information entropy:
\begin{equation}
  \label{eq: Shannon2}
  S_f(t) = - \sum_i^{\infty} p_i (t) \ln  p_i (t).
\end{equation}

\begin{figure}
\includegraphics[width=8cm]{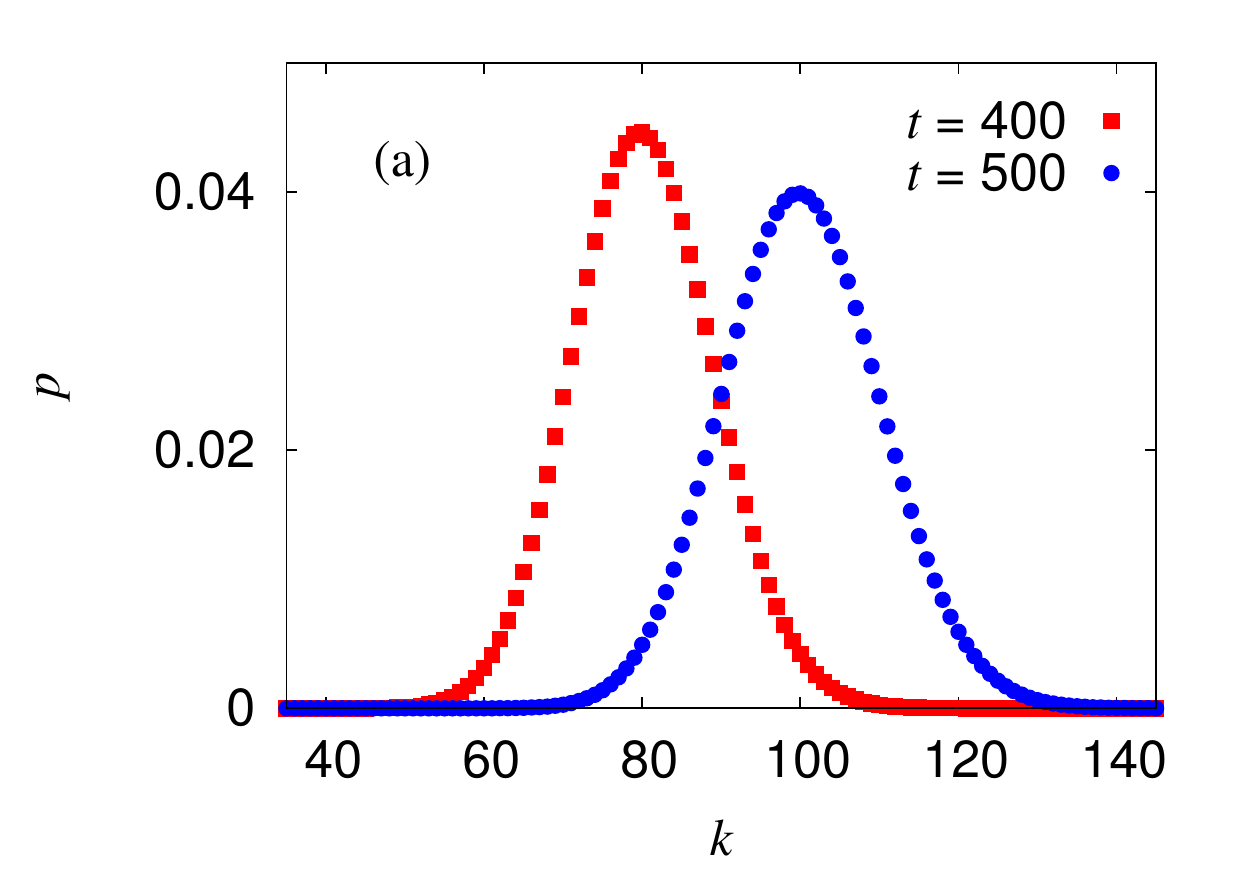}
\includegraphics[width=8cm]{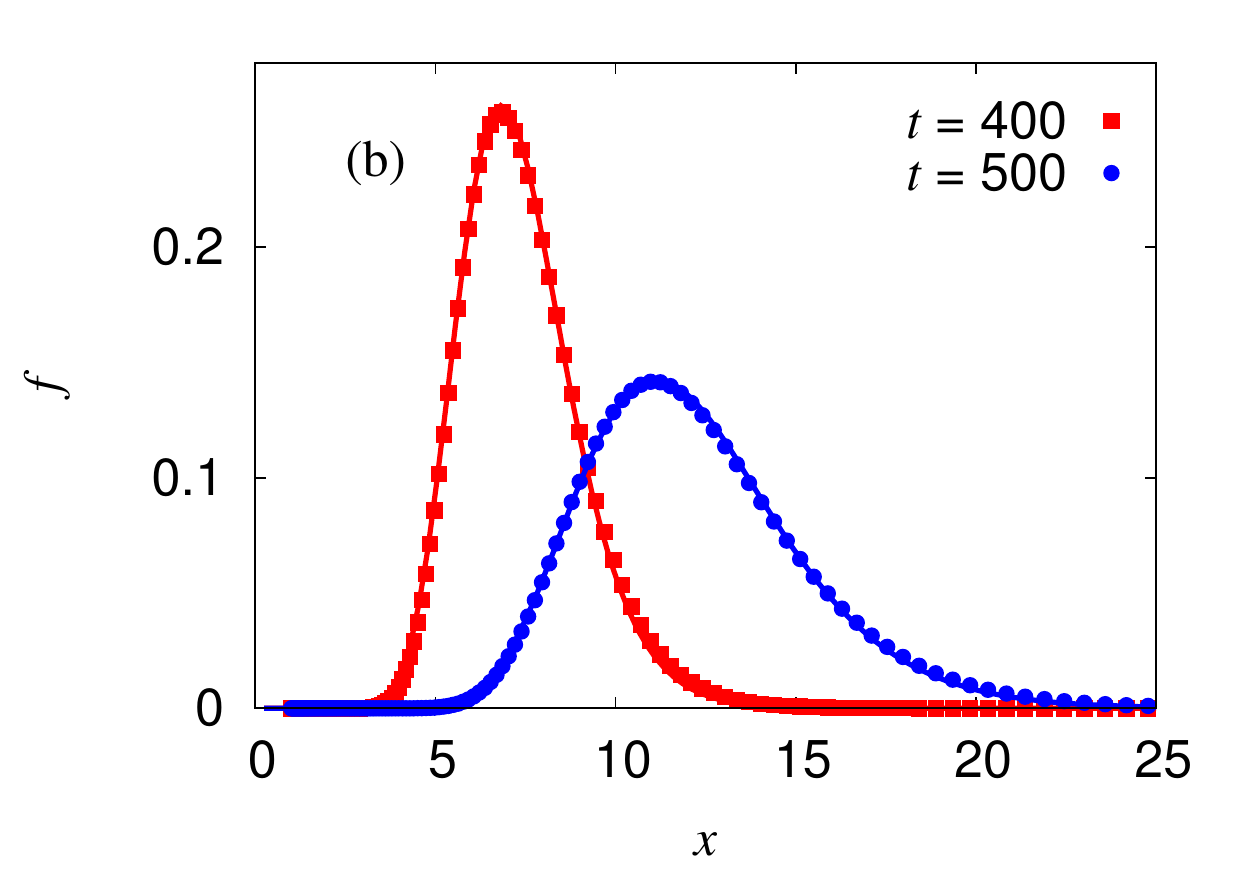}
\includegraphics[width=8cm]{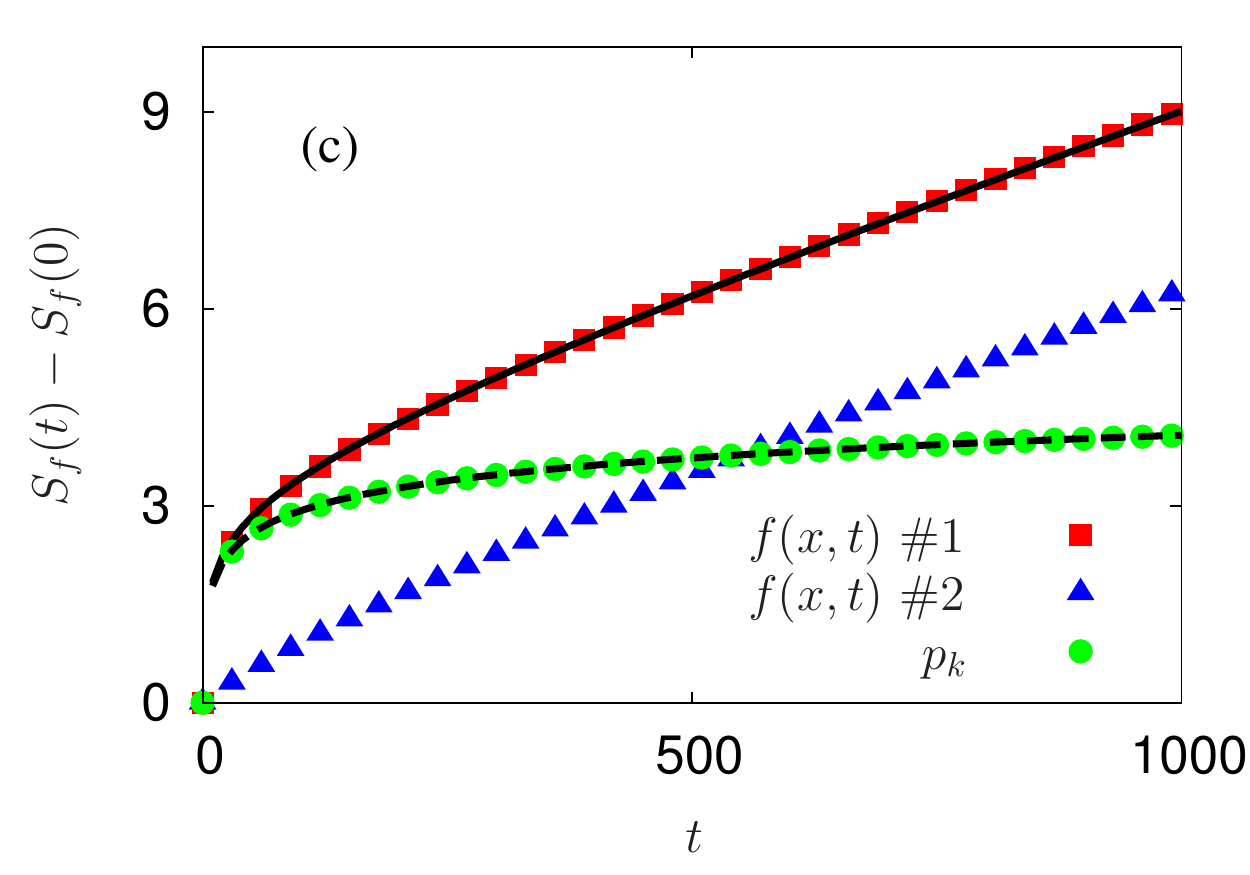}
\caption{(Color online) Size distributions and time evolutions of entropy in the number conserving system. In the continuous description, parameters $\lambda = 0.2$ and $b=0.025$ are used. (a) Discrete size distribution $p_k (t)$ versus $k$ and (b) (continuous) size distribution $f(x,t)$ versus $x$ at time $t=400$ (red squares) and $500$ (blue circles). Solid lines in (b) depict the log-normal distributions with $(\mu_t ,\sigma_t ) = (1.97, 0.218 )$ and $(2.47, 0.244)$, respectively. (c) Coarse-grained entropy $S_f(t)-S_f(0)$ versus time $t$. Red squares describe the entropy calculated from a narrow uniform initial distribution and blue triangles that from a Gaussian initial distribution. Green circles represent the entropy for the discrete size distribution. Black solid and dashed lines present the analytical results given by Eqs.~\eqref{eq: num conserved ent asympt} (for the continuous description) and ~\eqref{eq: num conserved ent asympt disc} (for the discrete description), respectively.
}
\label{fig: no production}
\end{figure}

Figure~\ref{fig: no production} presents the results for the number conserving system. In Fig.~\ref{fig: no production}(a) we display the discrete size distribution $p_k$ at time $t=400$ and $500$ in a system with  $\lambda = 0.2$ and $b=0.025$. As time goes by, the Poisson distribution in Eq.~(\ref{eq: poisson}) approaches the normal distribution peaked at $\lambda t$ with the standard deviation $\sqrt{\lambda t }$. The data in Fig.~\ref{fig: no production}(a) indeed fit well with these values of the peak position and the standard deviation (results not shown in the figure). Figure~\ref{fig: no production}(b) shows the size distribution $f(x,t)$ at time $t=400$ and $500$ for the system with the same model parameters $\lambda$ and $b$. The data in Fig.~\ref{fig: no production}(b) may be obtained from those of $p_k$ via the relation
\begin{equation}
\label{eq: p vs f}
     p_k (t) \delta k = f(x , t) \delta x
\end{equation}
with $\delta k =1$ and $\delta x = x [(1+b)^{1/2} -(1+b)^{-1/2}]$.
As addressed already, $f(x,t)$ reduces to the log-normal distribution in the long-time limit. Indeed, starting from the normal distribution
\begin{equation}
  \label{eq: normal distribution}
  p_k (t)  =\frac{1}{\sqrt{2 \pi \lambda t} }\exp \left[-\frac{ \left(k - \lambda t  \right) ^2}{2 \lambda t} \right]
\end{equation}
and putting $x=(1+b)^k$ with $k$ regarded as a continuous variable,
one can also obtain Eq.~(\ref{eq: log-normal distribution}) with $\mu_t = \lambda t \ln (1{+}b) $ and $\sigma_t = \sqrt{ \lambda t} \ln (1{+}b) $.
Fitting the data in Fig.~ \ref{fig: no production}(b) to Eq. (\ref{eq: log-normal distribution}), one finds excellent agreement with the theoretical values of $\mu_t$ and $\sigma_t$.

Figure~\ref{fig: no production}(c) shows the entropy, growing in time, for the same system. Red squares represent the entropy obtained from a narrow uniform initial distribution (labeled as \#1) via the simplified way of space integration described in Sec. II and blue triangles that from a Gaussian initial distribution (labeled as \#2). Green circles present the entropy for the discrete size distribution. Also shown are black solid and dashed lines representing the analytical results given by Eqs.~\eqref{eq: num conserved ent asympt} (for the continuous description) and ~\eqref{eq: num conserved ent asympt disc} (for the discrete description), respectively. Agreement between analytical solutions and numerical results is manifested. Note that the difference $S_f(t)-S_f(0)$ is displayed and the two data sets \#1 and \#2 fall in almost with each other eventually, except for the more rapid increase of the data set \#1 reflecting the lower entropy for the uniform distribution. If we compare the entropy of the continuous system designated by \#1 and that of the discrete system, the initial increases are similar but the latter grows more slowly. In particular, the increase becomes almost linear at large time $t$ and the slope computed analytically agree well with the numerical values, explaining the more rapid increase of $S_f(t)$ (and $S(t)$ as well) in Fig.~\ref{fig: no production}(c).

This rapid increase in the continuous description is attributed to the use of the domains of the real space growing exponentially in time, as confirmed easily by computing directly the difference.
From Eqs.~\eqref{eq: differ 2} and \eqref{eq: p vs f}, one obtains
\begin{align}
\label{eq: ent dk vs ds}
S_f(t)&=-\sum_{i}^\infty \frac{p_i (t)}{\delta x}\ln{\frac{p_i (t)}{\delta x}}\delta x \nn \\
&=-\sum_i^\infty p_i (t)\ln{p_i (t)} +\ln{(1{+}b)}\sum_{k=0}^\infty kp_k (t) \nn \\
&\quad +\ln{\left( \sqrt{1+b} +\frac{1}{\sqrt{1+b}} \right)}.
\end{align}
Neglecting the constant term in the asymptotic limit ($t\to \infty$), the difference is exactly given by the second term originating from the extension of the size space. This term turns out to be $\lambda t\ln{(1{+}b)}$, which confirms the analytical results given in Eqs.~\eqref{eq: num conserved ent asympt} and~\eqref{eq: num conserved ent asympt disc}. Apart from the constant shift, the numerical results are shown to fit well with the analytical results.

In general when the initial size distribution is very sharp, the size growing in time tends to take discrete values and rather a discrete size distribution is maintained in finite time, making the results for the discrete description applicable. On the other hand, if the initial distribution is somewhat broad, diversity of size is generated by the growth process and the continuous size distribution should be relevant. Another point to mention is that the success of the simplified method of integration for the differential entropy is related to the measurement scale on the element size. The logarithmic scale is thus more appropriate for the size in this growth problem.

\begin{figure*}
\includegraphics[width=8cm]{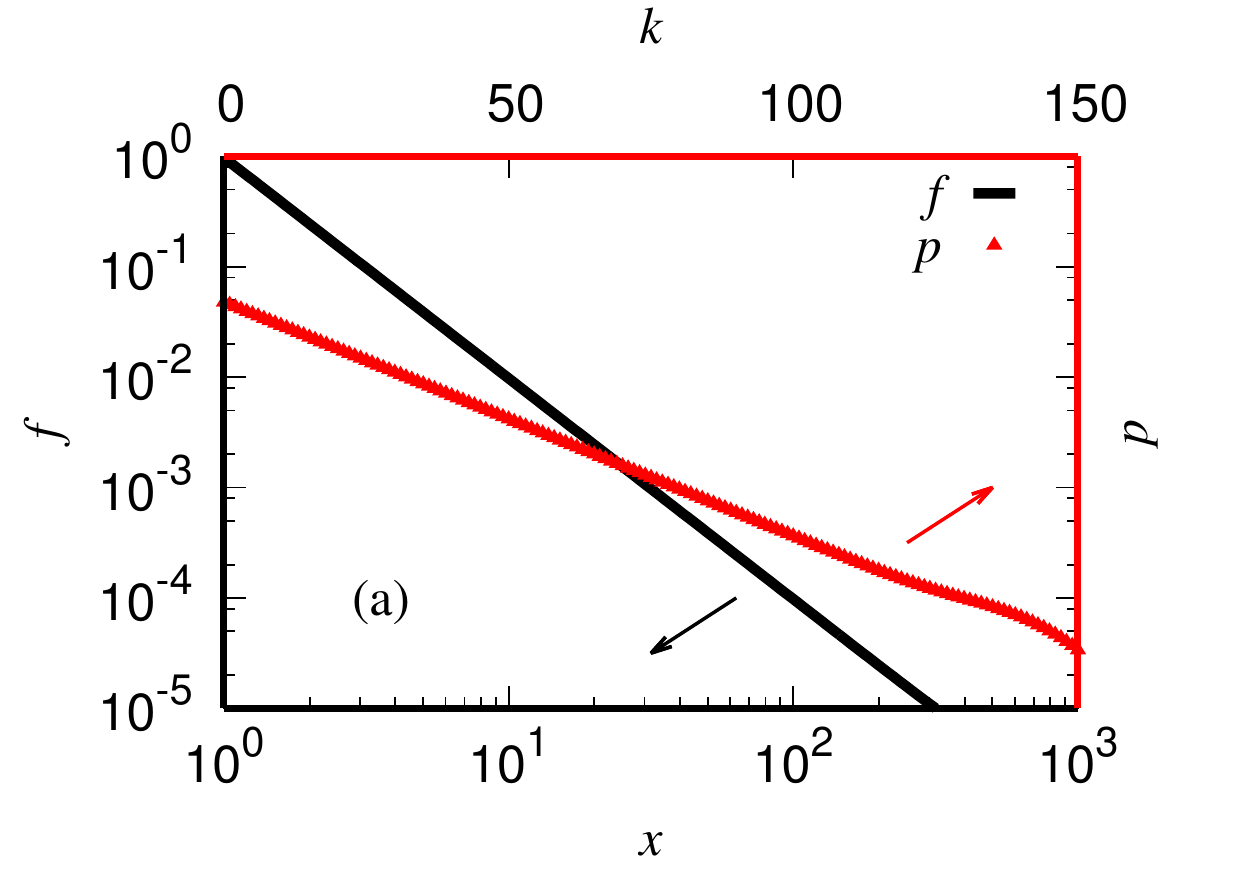}
\includegraphics[width=8cm]{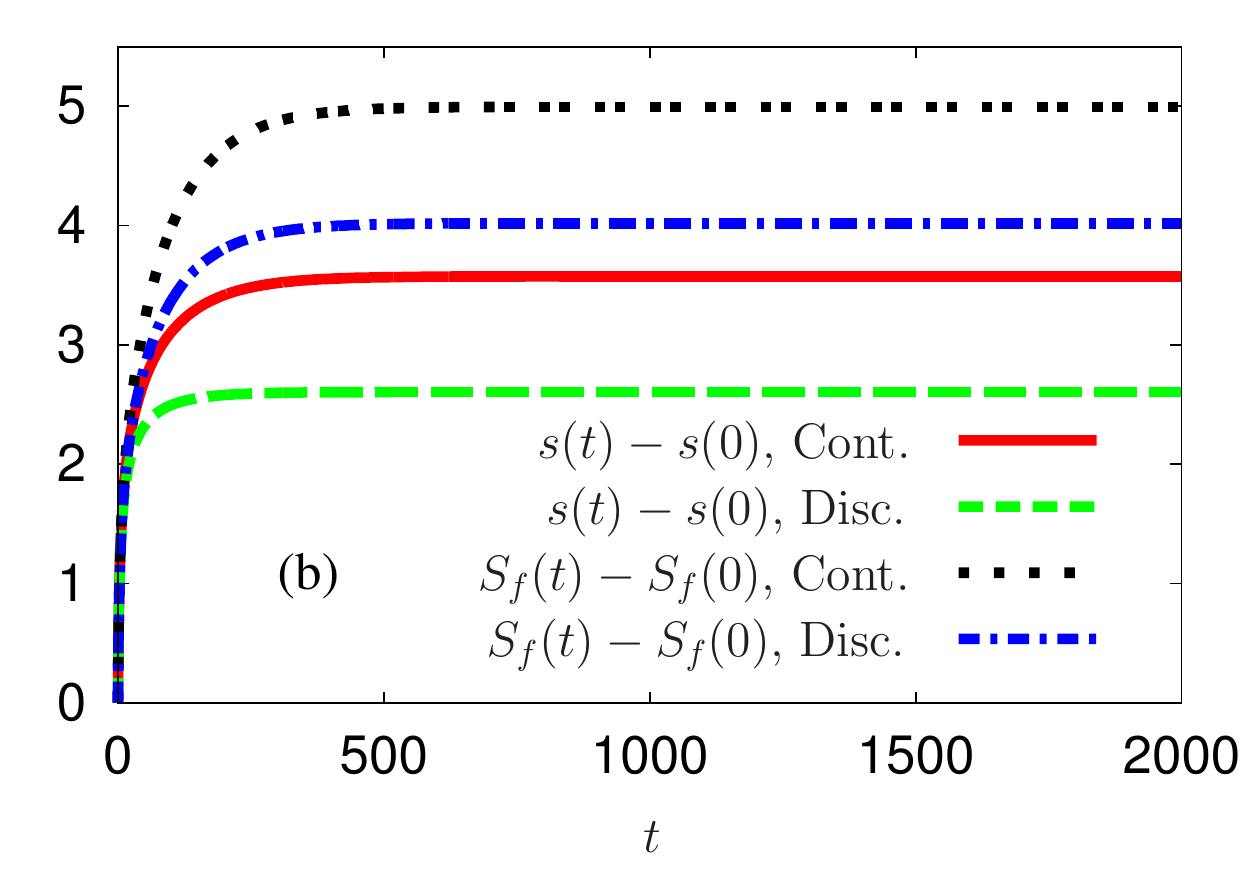}
\caption{(Color online) Size distributions and evolutions of entropy in the case of uniform size production for a system with $(r, \lambda) = (0.01, 0.2)$. In the continuous description, the growth factor is taken to be $b=0.05$. (a) Discrete and continuous size distributions $p_k (t)$ and $f(x,t)$ at time $t=700$.
Red triangles plot the data for $p_k$ in the semi-log scale while black solid line those for $f(x, t)$ in the log-log scale.
(b) Time evolution of entropy. Red solid, green dashed, black dotted, and blue dot-dashed lines correspond to the fine-grained entropy in the continuous description, fine-grained entropy in the discrete description, coarse-grained entropy in the continuous description, and coarse-grained entropy in the discrete description, respectively. In each case, a uniform initial distribution located narrowly at $x=1$ is used. It is manifested that the coarse-grained entropy is larger than the fine-grained entropy.
}
\label{fig: uniform}
\end{figure*}

\begin{figure*}
\includegraphics[width=8cm]{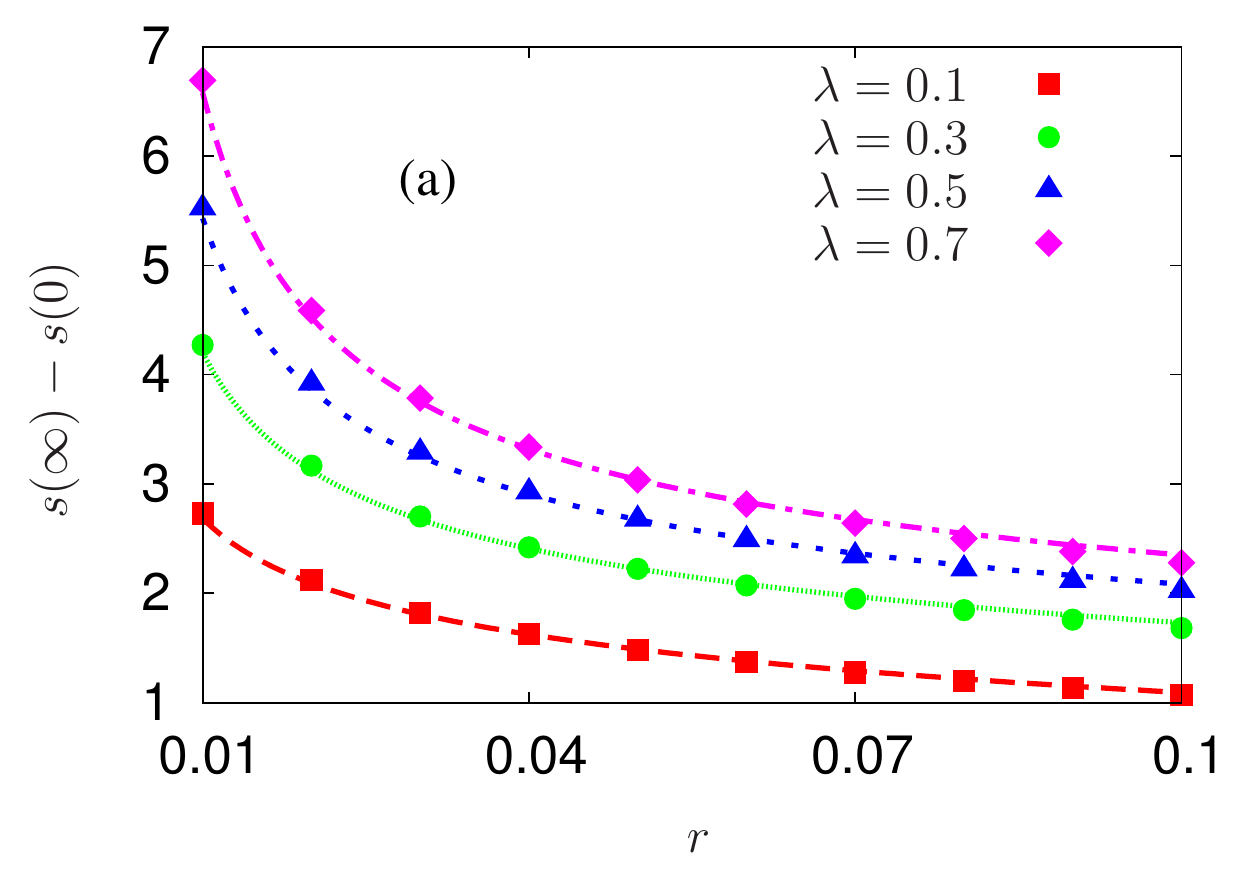}
\includegraphics[width=8cm]{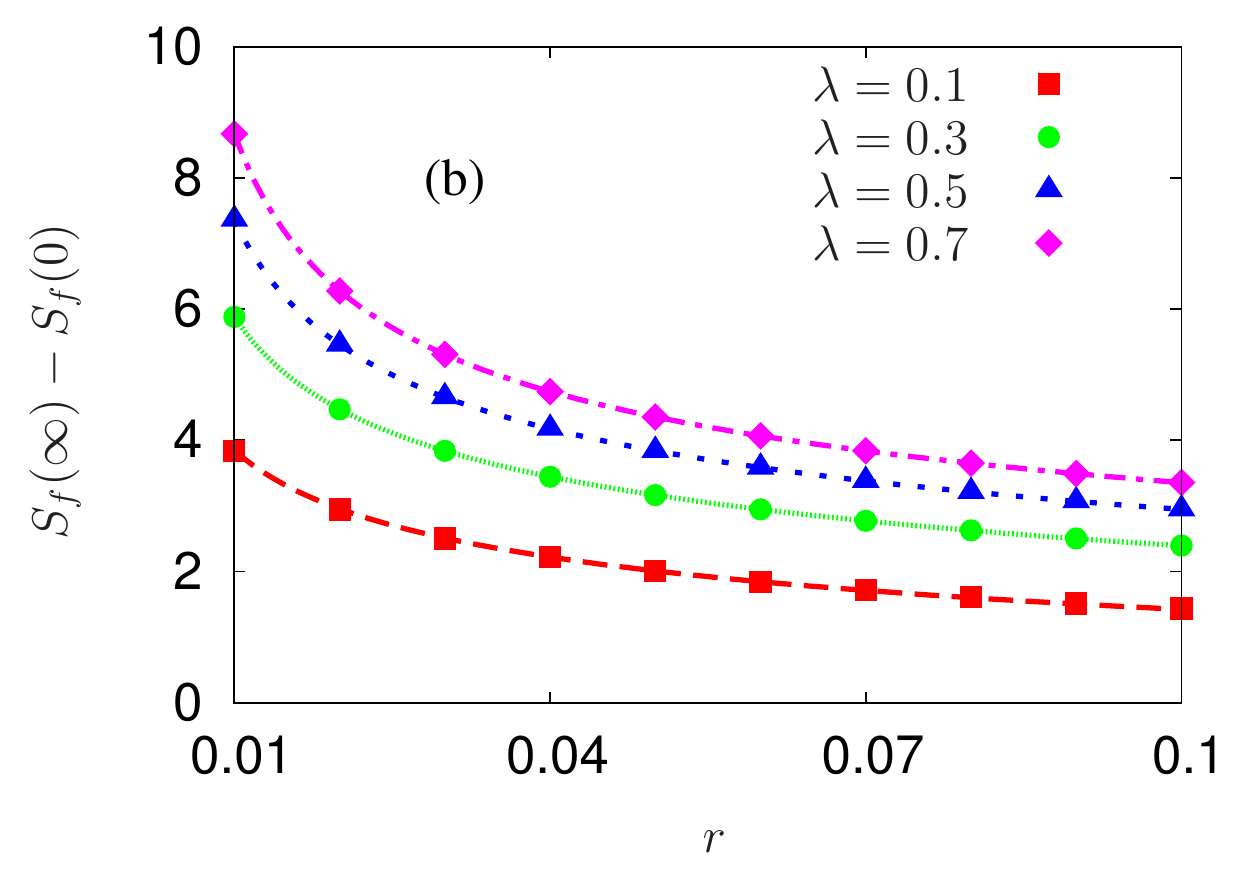}
\caption{(Color online) Stationary values of entropy in the case of uniform size production for $b=0.05$ and various values of $\lambda$ and $r$.
(a) Fine-grained entropy $s(\infty)-s(0)$ and (b) coarse-grained entropy $S_f (\infty)-S_f (0)$ versus $r$. Red squares, green circles, blue triangles, and pink diamonds plot the stationary values computed from numerical simulations for $\lambda=0.1, 0.3, 0.5$ and $0.7$, respectively. Red dashed, green solid, blue dotted, and pink dot-dashed lines represent the values given by (a) Eq.~\eqref{eq: fine h stationary} and (b) Eq.~\eqref{eq: coarsen h stationary} with $\lambda=0.1, 0.3, 0.5$ and $0.7$, respectively. The agreement between the analytical solutions and numerical results is evident in both (a) and (b). It is also observed that the coarse-grained entropy is always larger than the fine-grained entropy, which reflects the loss of information due to the disregard of heterogeneity during the coarse-graining procedure.
}
\label{fig: stationary}
\end{figure*}

In addition, we have noticed two kinds of entropy for the systems studied.
In the uniform production case, we compute the fine-grained entropy in addition to the coarse-grained entropy, to check whether or not the two are equivalent. In this case, we trace the birth of a new element together with the time of birth. From the age of each element at given time, we compute the contribution of each element to entropy. At the end, we sum the contributions over all elements and obtain the total entropy, making use of the extensiveness of entropy, and present the results of both fine-grained and coarse-grained entropies in Figs.~\ref{fig: uniform} and \ref{fig: stationary}.

Figure~\ref{fig: uniform}(a) shows the size distributions $p_k (t)$ and $f(x,t)$ for the system with parameters $(r, \lambda, b) = (0.01, 0.2, 0.05)$ at time $t=700$. Red triangles present the data for $p_k$ in the semi-log scale while (black) solid line plots those for $f(x, t)$ in the log-log scale. Both plots are observed linear for not too large values of $x$ and $k$. Here the linear region tends to expand with the lapse of time, and a stationary state is reached finally. In Fig.~\ref{fig: uniform}(b), we display how coarse-grained entropy as well as the fine-grained entropy evolves in time in the case of uniform-size production for the system of Fig.~\ref{fig: uniform}(a). It is shown that the entropy becomes nearly saturated after the power-law distribution is established and accordingly the stationary state is reached. It is evident that the coarse-grained entropy is larger that the fine-grained entropy, as expected from Eq.~\eqref{eq: S neq Sf}. Further, the entropy in the discrete description is smaller than that in the continuous description.

Finally, the stationary values of $s$ and $S_f$ for various values of $\lambda$ and $r$ are shown in Fig.~\ref{fig: stationary}(a) and (b), respectively (see the legend for the details). It is shown that the results fit very well with the analytical results. In particular, we observe that $s \neq S_f$ and it is confirmed that the coarse-grained entropy is larger than the fine-grained entropy. We thus conclude that the heterogeneity in addition to the correlations between elements can induce loss of information in the coarse-grained procedure.

\section{Summary}
We have studied the entropy of a system of elements evolving according to the master equation. Specifically, we consider the growth model in the fine-grained description, where the probability of the system configuration is governed by the master equation, and in the coarse-grained description, which deals with the evolution equation for the distribution function.
The system which accommodates production of new elements as well as the number conserving system without production have been probed in detail. Further, the difference between the two cases of the size variable, continuous and discrete (size) descriptions has also been examined.  What has been revealed and its implications are summarized in the following:

First, we have found that the growth rate of the size domain also provides an important factor for the time evolution of the entropy. Such growth of the domain is closely related to the resolution of description of the system. Indeed, in the discrete description, the domain is determined as the sum of the possible locations of elements in the size variable space, while the domain should span the whole space in the case of the continuous description. In some systems such as the classical random walk model, the volumes of the domains in the continuous and discrete descriptions are proportional to each other. In the case of the growth model studied in this paper, however, the resolution is directly connected with the size scale of the system via $\delta x \sim x$. Therefore, the domain increases faster as the size scale of the system grows larger, leading to the information loss due to the expansion of the domain space to be probed in the continuous description.

Second, examining the uniform production case of the growth model, we have confirmed that the heterogeneity in addition to the correlations among elements can induce loss of information or increase of entropy in the coarse-graining procedure. In this case, the entropy is still extensive but the coarse-graining procedure blurs out the disparity between elements (e.g., ages of produced elements) and as a consequence, causes the loss of information. The coarse-graining process, employed widely in the study of complex systems, may therefore yield a biased result unless heterogeneity is taken into account duly in the analysis. To quantify the amount of information loss accompanying the coarse-graining procedure should be very helpful for understanding the scale-dependent properties of complex systems. This is left for further study.

\section*{ACKNOWLEDGEMENTS}
This work was supported in part by the 2015 Research Fund of the University of Ulsan.

\end{document}